# Using Donald Trump's COVID-19 Vaccine Endorsement to Give Public Health a Shot in the Arm: A Large-Scale Ad Experiment[*]

**Bradley J. Larsen[§], Timothy J. Ryan[‡], Steven Greene[†],
Marc J. Hetherington[‡], Rahsaan Maxwell[‡], and Steven Tadelis[¶]**

March 29, 2022

**Abstract**

We report a large-scale randomized controlled trial designed to assess whether the partisan cue of a pro-vaccine message from Donald Trump would induce Americans to get COVID-19 vaccines. Our study involved presenting a 27-second advertisement to millions of U.S. YouTube users in October 2021. Results indicate that the campaign increased the number of vaccines in the average treated county by 103. Spread across 1,014 treated counties, the total effect of the campaign was an estimated increase of 104,036 vaccines. The campaign was cost-effective: with an overall budget of about $100,000, the cost to obtain an additional vaccine was about $1 or less.

[*] This research is approved under IRB-59690 (Stanford University). We benefited from financial support from the Vaccine Confidence Fund and from conversations with Cameron Ballard-Rosa, James Chu, Isabella de Vere Hunt, Pascaline Dupas, Matt Gentzkow, Paul Gertler, Guido Imbens, Randall Lewis, Eleni Linos, Mike MacKuen, Santiago Olivella, Linda Ong, Christopher Palmer, Kurt Ribisl, Jason Roberts, Darcy Sawatski, Hal Varian, and Robb Willer. We also benefitted from collaboration on related vaccine messaging efforts with the National Military Family Association and the Joseph Rainey Center for Public Policy.

[§] Stanford University Department of Economics, NBER, and SIEPR; bjlarsen@stanford.edu
[†] North Carolina State University Department of Political Science
[‡] The University of North Carolina at Chapel Hill Department of Political Science
[¶] University of California Berkeley Haas School of Business, NBER and CEPR

# 1.     Introduction

In June 2021, U.S. White House chief medical advisor Anthony Fauci warned that COVID-19 vaccination disparities could lead to the emergence of "two Americas," where regions with higher rates would fare significantly better than those with lower rates (*1*). This division would have a political bent. Republican political leaders often downplayed the virus's dangers, with some even amplifying false claims about the vaccine's safety, and vaccine hesitancy emerged disproportionately among Republicans. The Kaiser Family Foundation estimated that, among the 27 percent of American adults who remained unvaccinated in late October 2021, 60 percent were Republicans, far above their share in the electorate. Although race, ethnicity, income, urbanicity, education, and age were also associated with Americans' decisions to get vaccinated, political partisanship was the deepest fault line (*2*). A similar political divide exists along other CDC guidelines (*3, 4*). Consistent with Fauci's warning, the division carried life and death consequences. By early fall, counties that voted heavily for Donald Trump experienced COVID-related death rates nearly three times higher than counties that voted heavily for Joe Biden (*5*).

We posited that a remedy for this partisan divide might arise from mechanisms similar to those that created the disparity in the first place: partisan cues. Research shows that partisans form preferences by following cues from their party leaders (*6, 7*), a regularity that has grown stronger as the parties have polarized over the last generation (*8*). We hypothesized that messages publicizing Trump's support for COVID-19 vaccines—support Trump did little to advertise after leaving the White House—might cue the vaccine-hesitant among his supporters to get vaccinated themselves. To test this hypothesis, we created a public service announcement (PSA), featuring news clips of Donald Trump encouraging his supporters to get vaccinated. We tested its efficacy through a large randomized controlled trial (RCT) on YouTube, randomizing at the county level and targeting areas that lagged in vaccine uptake. Overall, we spent approximately $100,000 U.S. in ads, with a total of 11.6 million ads reaching 6 million unique viewers.

We measure the effect of the campaign on county-level vaccination counts in CDC data. We estimate that the treatment led to an increase of 103 vaccinations in the average treated county. This effect is significant at the 0.05 level in a two-tailed test under our pre-registered inference plan, which involves state-level clustering, and at the 0.20 level under county-level clustering. With 1,014 treated counties in all, the total increase in vaccinations from the campaign was 104,036, costing about $1 or less in ad spending per vaccine. This is a fraction of the cost of other interventions, such as vaccine

lotteries or interventions offering direct compensation for getting vaccinated (*9, 10, 11*). In short, we find that a problem with political origins also has a political remedy.

## 2. Ad Campaign Setup

### 2.1 Message Considerations

Previous research has demonstrated that message persuasiveness rarely hinges on argument quality, because evaluating quality requires effortful information processing that many people avoid (*12, 13*). Instead, people tend to rely on mental shortcuts, such as beliefs about a source's credibility, to decide whether to accept or reject new information (*14*). Original survey data we collected in the months following the pandemic's onset revealed that members of our target audience (unvaccinated Republicans) persistently had greater confidence in vaccine advice coming from Donald Trump than in advice coming from more traditional sources, such as their personal doctor or the scientific community (Table 1). Donald Trump has an additional advantage: because Republicans have become identified with skepticism about the severity of the COVID-19 virus and COVID vaccines, a pro-vaccine message from the leader of the Republican Party qualifies as counter-stereotypical—and counter-stereotypical messages have been shown to evoke more effortful mental processing (*15*). Indeed, one recent survey experiment conducted on a convenience sample found that a vaccine message from Donald Trump successfully increased vaccine *intentions* among Republicans (*7, 16*). We build on those insights but shift focus from intentions to actual vaccination *behavior*.

A pro-vaccine endorsement from Donald Trump that is associated with Fox News should further enhance the counter-stereotypical nature of the message. Fox News personalities have been skeptical of Fauci and many COVID vaccine efforts. Moreover, previous research has demonstrated the persuasive effects of Fox News on Republican attitudes broadly, as well as on viewers' attitudes toward pandemic guidelines in particular (*17, 18*).

### 2.2 The Public Service Announcement

The PSA includes four separate video clips – the first and third from a Fox 13 News Utah (a local station) segment recorded on March 16, 2021, the second from a phone interview between Donald Trump and anchor Maria Bartiromo recorded on the Fox News Channel (the nationwide cable TV channel) from the same date, and the fourth from a social media post of Ivanka Trump from the spring of 2021. We hired a professional video editor to combine these clips and overlay them with an exciting soundtrack.

Knowing that many users might opt to stop the PSA from playing as soon as possible, it was imperative that news of Trump's endorsement occur immediately. Within the first three seconds of our ad, the Fox 13 Utah anchor says, "Donald Trump is urging all Americans to get the COVID-19 vaccine." The rest of the PSA unfolds as follows:

- **Seconds 4–12:** Donald Trump speaking on phone interview on Fox News with Bartiromo, while Bartiromo nods in agreement: "I would—I would recommend it, and I would recommend it to a lot of people that don't want to get it, and a lot of those people voted for me, frankly."
- **Seconds 13–19:** Fox 13 News Utah anchor, with footage of the Trumps at White House: "Both Trump and former First Lady Melania Trump did receive their vaccines privately in January at the White House."
- **Seconds 20–23:** Screen text quote from Ivanka Trump, with still shot of her getting the vaccine: "Today I got the shot. I hope you do too."
- **Seconds 24–27:** Black screen with white print: "Your vaccine is waiting for you."

The specific YouTube setting we adopted required users to watch the first five seconds of the PSA before being allowed to skip. Hence all viewers, even if only involuntarily, heard that Trump was urging all Americans to get the vaccine. A screenshot taken at 2 seconds into the ad appears in Figure 1. One second later, viewers saw Fox News' familiar news anchor, Bartiromo, and the official Fox News stamp. For viewers who had the sound on their devices muted, on screen text delivered the message. The Fox Utah news story prominently displays, "Trump encourages getting COVID-19 vaccine" on the screen, and the Fox News story shows, "Trump on the success of operation warp speed." We also included closed captions of all spoken words to increase the likelihood that viewers would absorb the message. We also embedded a link in the bottom left corner of the ad, allowing viewers to click to see the full Fox News interview with Bartiromo. In the bottom right, a box displays a countdown of the number of seconds until the user can skip the ad (showing "3" in Figure 1). The PSA can be viewed at https://www.youtube.com/watch?v=INH-CmCgIYs.

### 2.3 How YouTube's Advertising Platform Works

The YouTube advertising platform (Google Ads) allowed us to select a target population (our treatment counties) and to specify our willingness to pay for 1,000 ad impressions, known as a "mille" in advertising lingo. Below, we refer to the Cost Per Mille (CPM). For each advertising slot—someone watching a YouTube video—Google runs an instantaneous auction, bidding on behalf of each advertiser, to determine which advertiser's content will be shown. An advertiser is more likely to win an auction if she is willing to pay more than other advertisers and if Google predicts that, for a given user, the advertiser's content is more likely to generate user engagement (which Google defines as watching at least ten seconds of the ad). As we describe below, these features generated widely varying exposure to our ad across different counties, as Google's algorithm dynamically adjusted as it learned which types of users were more likely to engage with the ad.

### 2.4 Selection of Treatment and Control Counties

We designed our experiment to concentrate ads in areas with low vaccination rates (below 50%), moderate population sizes (fewer than 1 million residents), and well-recorded CDC vaccine data. Our supplemental materials provide additional details. These exclusions left us with 2,168 counties, which we randomly assigned to treatment and control in a stratified manner to maintain balance on population and vaccination status. This procedure resulted in 1,083 counties assigned to receive ads, and 1,085 retained as a control group. Treatment and control groups were closely balanced on prior percentage of the population vaccinated ($\text{diff} = 0.01, SE = 0.39, p = 0.98$) as well as population ($\text{diff} = -3,216, s.e. = 3,757, p = 0.39$). Among these counties, 136 (69 treatment and 67 control) ended up reporting no vaccine count data to the CDC during our sample period. Our regression analyses below therefore use 1,014 treatment counties and 1,018 control counties, spanning 43 states.

### 3. Ad Campaign Characteristics

Online advertising often results in low signal-to-noise ratios, requiring very large ad campaigns to detect an effect even in carefully designed randomized control trials (*19, 20*). Fortunately, our ad budget was substantial: $99,009.51 total, $96,408.56 of which was spent in counties that reported vaccine data to the CDC. The campaign ran from October 14 through October 31, with ads steadily rolling out over this period, as shown in Figure 2.A. We initially avoided any targeting of specific demographic groups other than excluding minors. After the first 10 days of the campaign, however, we observed in Google Ads tools that users ages 18-24 were receiving a disproportionate share of ads. In response, we excluded this age category for the last week of the campaign. Combined, we purchased a total 11,573,574 impressions, with an average CPM of $8.55. These ads were delivered to 6,079,732 distinct viewers, with the average user seeing the ad 1.9 times and no one seeing it more than four times.

Google Ads tools allow us to observe characteristics of YouTube viewers and the outlets through which they viewed the ad. 52% of ads appeared on phones, 30% on television screens (e.g., via Roku or Apple TV), 13% on tablets and 4% on computers. We find the viewers watching on TV screens tended to watch much more of the ad. The PSA was placed on 150,284 distinct YouTube channels, several hundred websites, and 10,072 mobile apps (largely games). Figure 2.B shows that, of these YouTube channels, the Fox News channel by far hosted our ad the most—over 200,000 times, which is three times the quantity on Forbes and ten times that on NBC News YouTube channels. Google Ads tools also show the specific YouTube *postings* to which our ad was attached for a given channel. For the Fox News channel, these include segments by cable news personalities such as Laura Ingraham, Greg Gutfeld, Tucker Carlson, Judge Jeanine Pirro, Sean Hannity, and The Five. The ad also appeared on other channels with conservative leanings, such as Glenn Beck and The Blaze, as well as outlets not supportive of Trump, such as Saturday Night Live (each of these channels had over 3,000 ad impressions), and many local news stations. We emphasize here that the outlet was not a choice variable in our design; Google's algorithm *chose* to place our ad on these YouTube channels based on its predictions of user engagement, a highly convenient feature for our goal of reaching a vaccine-hesitant audience using a counter-stereotypical messenger.

## 4. The Overall Effect of the YouTube Ad Campaign

We now describe our empirical approach and results. We pre-registered our analysis plan via the Open Science Framework (see Supplemental Materials). Our pre-registered dependent variable is the cumulative number of COVID-19 vaccine first doses administered in each county up through a particular date. We focus only on vaccine recipients who are 12 or older, as children ages 5-11 only became eligible after our campaign ended (in early November 2021). An observation in our analysis is a given county on a given date. Our sample period encompasses dates from one month before the campaign to one month after, which includes 151,945 county-date observations.

We first analyze the overall effect of the campaign on the average treatment county, a county assigned to the treatment group by our random assignment algorithm. Our approach relies on *difference-in-difference* regressions estimated via ordinary least squares (OLS). These regressions compare the difference in the vaccine count between treatment counties and control counties *before* the campaign started to this same difference *after* the campaign. The regressions control for county-level differences in vaccine counts through county fixed effects, differences over time through date fixed effects, and differences in the trend of vaccine count growth rates over time in large vs. small counties. This third set of controls represents a departure from our pre-registered regression analysis plan. We discovered the need to control for differential growth rates by county population only after the campaign was complete. The Supplementary Materials discuss this and, for transparency, report our pre-registered specifications.

Because of how YouTube delivers ads (see Section 2.3), not all treatment counties received ads, and for those that did receive ads the exposure varied widely across counties, as we describe in greater detail in Section 5 below. Thus, the effect we measure in this analysis is the overall effect of the campaign on a county that is *assigned* to be a treatment county, regardless of the actual level of ad exposure. This is what is referred to as the *intent-to-treat* effect. For inference, as described in our pre-registration plan, we estimate standard errors clustered at the state level to allow for correlated residuals within states over time. This typically yields larger standard errors than county-level clustering. However, there are arguments in the methodological literature for reporting county-level clustering instead, and in our analysis, we found that county-level clustering yields larger standard errors. We therefore report standard errors under both types of clustering for comparison. The Supplemental Materials contain details.

The results of this analysis are shown in columns 1-2 of Table 2. Column 1 controls for differential trends by county population size through an interaction of a county population with a dummy for the period after the start of the campaign. Column 2 includes instead more flexible controls for these trends that interact county population size with separate dummy variables for individual dates. The results are statistically indistinguishable in the two columns, demonstrating that the results are not sensitive to how we control for the growth rate of vaccinations in different sized counties. In column 1, we observe an increase of 102.6 vaccines (95% CI 9.3 to 195.9 when clustered at the state level, and from –51.7 to 256.9 when clustered at the county level). In column 2 we find a similar positive effect of 101.4.

The estimate in column 1 implies that, among the 1,014 treatment counties, the average county saw a causal increase of 102.6 vaccines due to the campaign. Aggregating across all treatment counties, the total effect was an increase of 104,036 vaccines. Because our total ad budget spent in these counties was $96,408.56, the total cost in ads *per vaccine* was $0.93—less than $1 in advertising spending yielded one more shot in the arm.

## 5. The Effect of the Degree of Ad Exposure by County on Vaccinations

Our pre-registered research design focused only on a comparison between treated and untreated counties. However, the ad auction described in Section 2.3 introduced considerable variation in the number of ads received per county. This opens another axis of comparison, as effects may vary depending on the number of ads received. We analyze these effects here.

### 5.1 Variation in Ad Exposure Across Counties

Google Ads metrics records the number of ad impressions displayed in each county over the duration of the campaign. Because of the ad auction features described in Section 2.3, our campaign's budget was spent asymmetrically across our treated counties. Figure 3.A shows a histogram of the number of ads received by each county, omitting the top 5% for readability. Some counties received more than 40,000 ads (with a maximum of 346,089), while some received far fewer (including five small, treatment-assigned counties that received zero). This is because YouTube does not treat counties as separate blocks, each to be assigned some ad exposure. Instead, the whole list of counties is taken to be a single target audience and YouTube attempts to serve the ad to the users within that audience who are most likely to engage with the ad. The average county saw 10,679 ads, with a standard deviation of 25,245.

Figure 3.B normalizes the number of ads by the county population, showing a histogram of the number of ads per 100 residents in the county separately for large (those with a population above the median) vs. small counties, omitting the top 5%. These numbers range from 0 to nearly 30 ads per 100 residents for both sets of counties (with a maximum of 160), demonstrating that, even accounting for population differences, the number of ads is quite variable across counties, with the number of ads per capita being higher in larger counties. Figure 4 shows that this variation has no obvious geographical bias: high- and low-saturation counties are distributed more or less evenly throughout the entire United States.

**5.2 Measuring the Effect of an Increase in Ads**

We now exploit the variation described in Section 5.1, moving beyond the intent-to-treat effect to analyze how the *number of ads* a county receives affects the number of vaccines in the county, referred to as the *average causal response* (*21*). To measure this effect, we implement the instrumental variables (IV) design proposed by Angrist and Imbens in *(21)*, instrumenting for the number of ads in each county using the county's random assignment to treatment or control.

The reason for this design is important: unlike assignment to treatment or control status, the number of ads a county receives is not randomly assigned, but rather arises from the black box of Google's machine learning predictions of viewers' likelihood of engaging with the ad. As such, it is possible that Google sends more ads to counties where viewers are more likely to be receptive, and hence a standard OLS regression treating the number of ads as randomly assigned would not yield an unbiased estimate of the effect of ad exposure. (This does not bias the estimate of intent-to-treat effect, only the naïve estimate of the causal effect of an increase in ad exposure.) The IV approach, on the other hand, exploits the random assignment to restore a causal estimate of the response to an increase in the number of ads. Our setting also falls into the special case of *one-sided noncompliance*, meaning here that some treatment counties received no ads, but no control counties received ads (*22*). In this case, our effect corresponds to the average causal response *for treated counties* from an additional 1,000 ad impressions.

We display estimates from this IV regression in columns 3 and 4 of Table 2, which offer two different approaches to controlling for differential trends by county population, as in columns 1 and 2. Both columns 3 and 4 show a positive and significant (and similarly sized) effect. The estimate in column 3 of Table 2 suggests that an increase of 1,000 ads leads to 8.6 additional vaccines in the average treated county. Put differently, 116 ad impressions (i.e. 1000 divided by 8.6) are required to yield one additional vaccine. Because the standard deviation of the number of ads across treatment counties is 25,245, the estimate in column 3 implies that a one-standard deviation increase in ads yields 217 additional vaccines in the county.

As described in Section 3, 1,000 ad impressions cost us $8.55 on average (our average CPM). These numbers therefore imply a second estimate of the cost per vaccine, at $0.99. The estimate in column 4 implies a similar cost per vaccine, at $1.01. Thus, regardless of whether we rely on the estimated intent-to-treat effect or the average causal response, the advertising spending required to persuade one more viewer to get the vaccine is approximately $1 or less.

**5.3 Other Measures of Treatment Intensity**

The number of ads a county receives is one of several different ways of measuring a county's *treatment intensity*. Using the same IV framework from Section 5.2, we examine several other measures that represent viewer engagement. We estimate that a one-standard-deviation increase in the percentage of viewers watching at least ten seconds of the ad increases vaccines distributed in the average county by 8.2. Additionally, a one-standard-deviation increase in the rate of those viewing the full ad increases vaccinations by 12.3. Both suggest that counties in which viewers watched the ad for longer were the most responsive in terms of vaccinations. Other measures of viewer engagement produce even larger effects. A one-standard-deviation increase in the rate of clicking on the Fox News link at the bottom of the ad leads to 94.1 additional vaccines. These results are also reflected in the cost of ads: a one-standard-deviation increase in a county's CPM results in 4.9 additional vaccines, suggesting that, while we had to bid more for each ad in some counties than others, this increased spending yielded viewers who more likely to respond by getting the vaccine. Finally, ad coverage affects uptake as well, with a one-standard-deviation increase in the number of ads per capita in the county resulting in 48.4 additional vaccines. All these effects are statistically significant at the 0.05 level under state-level clustering and at the 0.20 level under county-level clustering. See Supplemental Materials for the full set of results.

## 6. Additional Analyses

In our supplemental material we show that the uptake of vaccines in response to the ad campaign was not immediate, but rather occurred later in the campaign, as ad exposure reached its peak, and in the weeks that followed. We also examine whether different types of counties responded differently to the ad treatment. We find no evidence that counties of different education levels or racial diversity responded differently to the campaign, but we do find a differential response depending on the share of Trump voters in a county. In particular, the ITT effect is positive and significant for counties in which fewer than 70% of voters voted for Trump, but for counties with more extreme Trump vote shares we cannot reject the null hypothesis that the campaign had no effect, suggesting that extremely right-leaning counties were less likely to be swayed by the campaign.

# 7. Concluding Remarks

The COVID-19 vaccines are a modern medical marvel. Developed in record time, they proved breathtakingly effective in reducing the chances of hospitalization and death among those who took them. Remarkably, a big obstacle to their adoption and use proved to be sociological: large segments of the population declined to be vaccinated, with partisanship exerting the largest effect. This resistance comes with negative externalities: more hesitancy increases the prevalence of "breakthrough" infections as well as the risk to immunocompromised individuals vulnerable to COVID infections even if vaccinated. Hesitancy in the U.S. proved especially high, with rates of vaccination barely among the top 50 countries in February 2022 (*23*). As many have observed, a tragedy of the COVID-19 pandemic is the extent to which protective measures became tangled in Americans' political identities, which led to deaths and suffering that could have been avoided.

But if politics characterizes one aspect of the problem, it might also point to part of a solution. We find that positioning Donald Trump as a counterintuitive messenger is a cost-effective way to overcome hesitancy among people who still had not been vaccinated, months after the vaccines became widely available.

One question that might arise about our findings is the extent to which they are scalable. We believe that, as long as Americans on the political right are a significant bastion of hesitancy, Donald Trump's support for vaccination will represent a potent tool that public health messengers can use. Leveraging his recent endorsement of COVID-19 boosters is an obvious extension of this work. So, too, with seasonal flu vaccines, which are starting to show signs of increased partisan schism (*24*). Insofar as vaccinations continue to be politicized, this research provides a model for political messaging as an important public health tool in future pandemics.

Whether the dividing line is politics or something else, we encourage public health proponents to reflect on messengers whose voices might carry special weight among target populations. For example, other research finds a pro-masking message from a military general increases support for masking among political conservatives (*25*). The results we report herein help corroborate this theme. We think it represents a promising route to overcome resistance and, in turn, save lives.

Table 1: Percent of Unvaccinated Republicans Expressing a
"Great Deal of Confidence" "when it comes to advising you on taking the COVID-19 vaccine."

|  | March/April | August/September |
|---|---|---|
| Donald Trump | 39 | 40 |
| Joe Biden | 10 | 8 |
| Anthony Fauci | 14 | 11 |
| Your Personal Doctor | 35 | 23 |
| Scientific Community | 16 | 11 |

Notes: Cell entries represent the percentage of respondents in a national survey indicating they have a great deal of confidence in vaccine advice coming from each of the indicated sources. This table relies on two different waves of the survey, March/April (sample size 422) and August/September (sample size 387). See supplementary materials for details on question wording and the sampling approach.

Figure 1: Screenshot of PSA at 3 Seconds

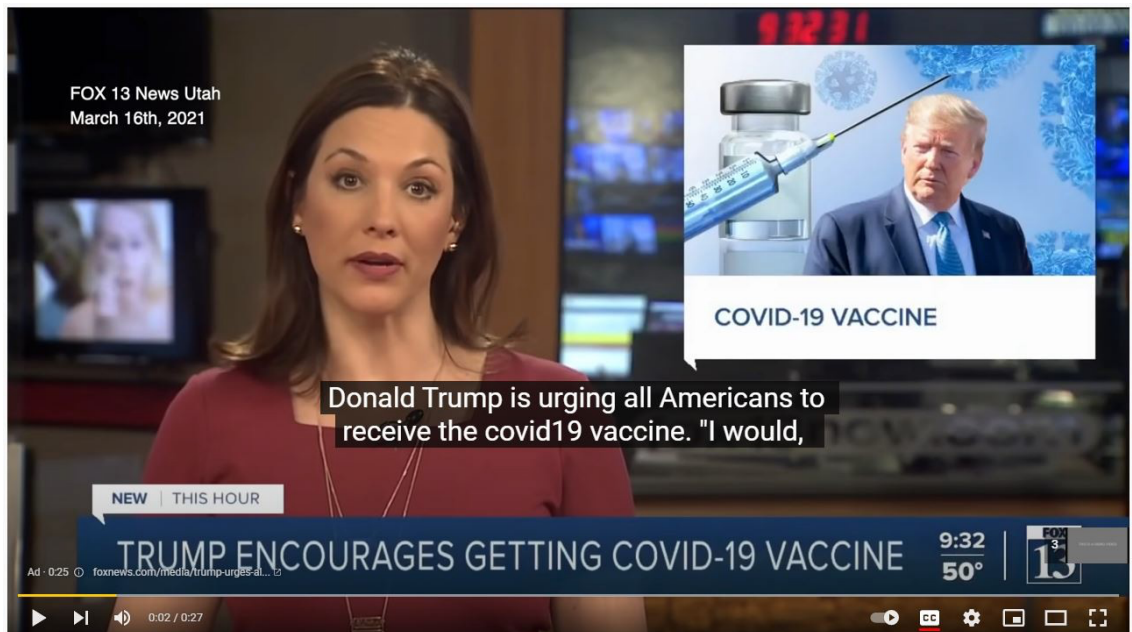

Notes: Closed captions by default appear towards the bottom middle of the screen, covering part of the "Trump encourages getting COVID-19 vaccine" text. The user can move these captions to different locations on the screen, and we did here to improve readability.

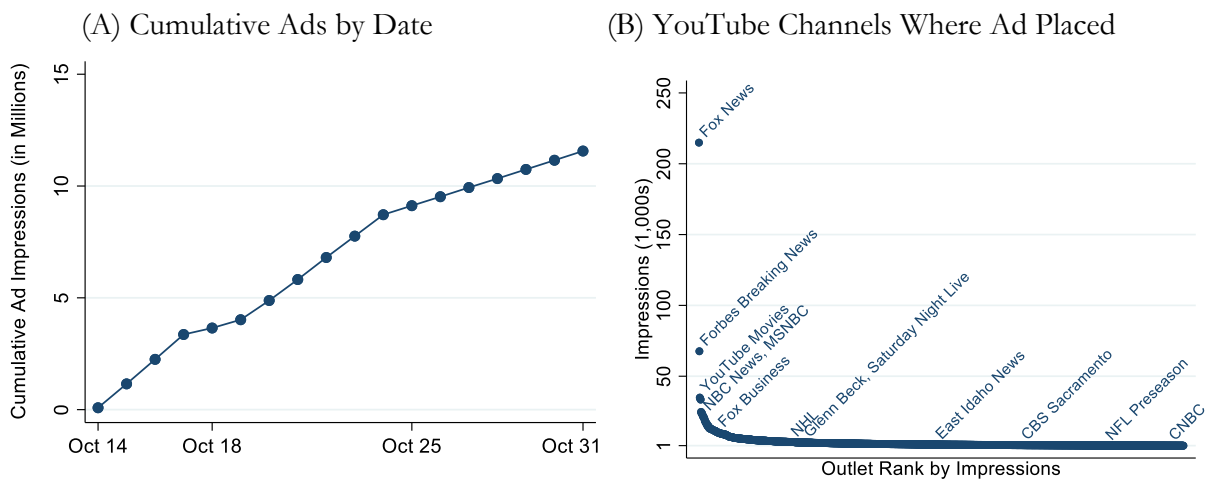

Figure 2: Frequency of Ads Across Counties

(A) Cumulative Ads by Date

(B) YouTube Channels Where Ad Placed

Notes: Panel A shows the cumulative ads by date (in millions). In Panel B, each dot represents a given YouTube channel on which the ad was shown over 1,000 times (with only a handful of these channels labeled for readability). The vertical axis shows the number of times (in units of 1,000) that the ad showed on a given channel, and the horizontal axis orders channels in descending order by the number of times the ad showed.

Table 2: Vaccine Increase Per County: Average Effect and Causal Response

|  | Intent-to-Treat Effect | | Average Causal Response of 1,000 Ads | |
|---|---|---|---|---|
|  | (1) | (2) | (3) | (4) |
| Effect | 102.6 | 101.4 | 8.606 | 8.500 |
|  | (46.65)** | (46.72)** | (3.986)** | (3.992)** |
|  | (78.74)+ | (78.76)+ | (6.608)+ | (6.609)+ |
|  |  |  |  |  |
| Implied Cost per Vaccine | $0.93 | $0.94 | $0.99 | $1.01 |
| County Fixed Effects | Yes | Yes | Yes | Yes |
| Date Fixed Effects | Yes | Yes | Yes | Yes |
| Pop. × Post Dummy | Yes |  | Yes |  |
| Pop. × Date Dummies |  | Yes |  | Yes |

Notes: Regression results. Sample size is 151,945 county-date observations. All regressions include fixed effects at the county and date levels. In columns 1-2, the intent-to-treat effect corresponds to the OLS-estimated coefficient on the interaction of a treatment assignment dummy (*Treat*) with a dummy for dates after October 14 (*Post*), the start of the campaign. Column 1 also includes the interaction of *Post* with county population. Column 2 replaces this with interactions of county population with (i) flexible dummies for each date within two weeks before to two weeks after the campaign (omitting the date before the campaign started), (ii) a dummy variable for two weeks or more before, and (iii) a dummy variable for two weeks or more after. Columns 3-4 report the IV-estimated coefficient on the interaction of the number of ads the county received (in 1,000s) with *Post*, with this interaction instrumented for by *Treat x Post*. Column 3 mimics column 1 in controlling for differential trends by population and column 4 mimics column 2. "***", "**", "*", and "+" indicate significance (from a two-tailed test) at the 0.01, 0.05, 0.10, and 0.20 levels. Standard errors, reported in parentheses below each estimate, are clustered at the state level in the first row and at the county level in the second row.

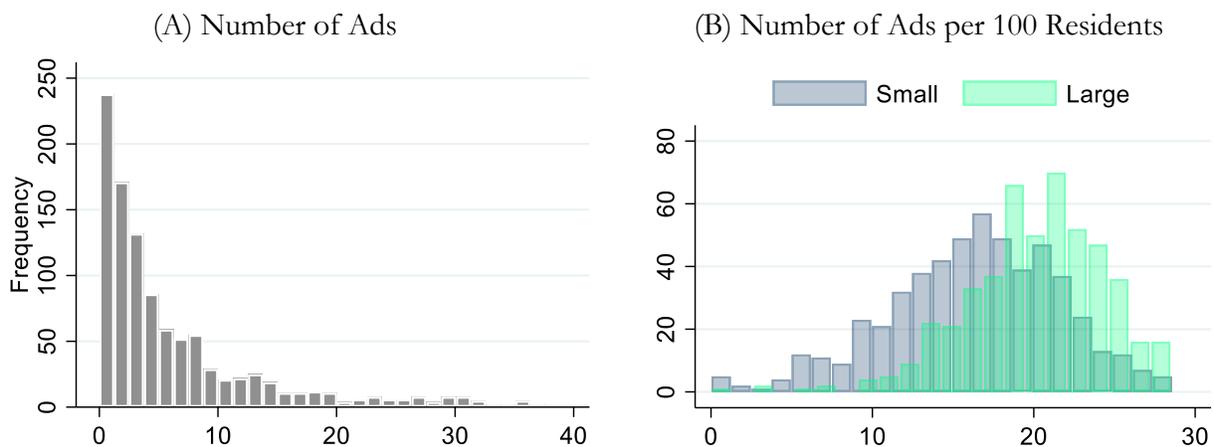

Figure 3: Frequency of Ads Across Counties

(A) Number of Ads  (B) Number of Ads per 100 Residents

Notes: Panel A displays histogram of the number of ads per county (in thousands). Panel B shows the number of ads per 100 county residents, separately for small (below median population) and large (above median population) counties.

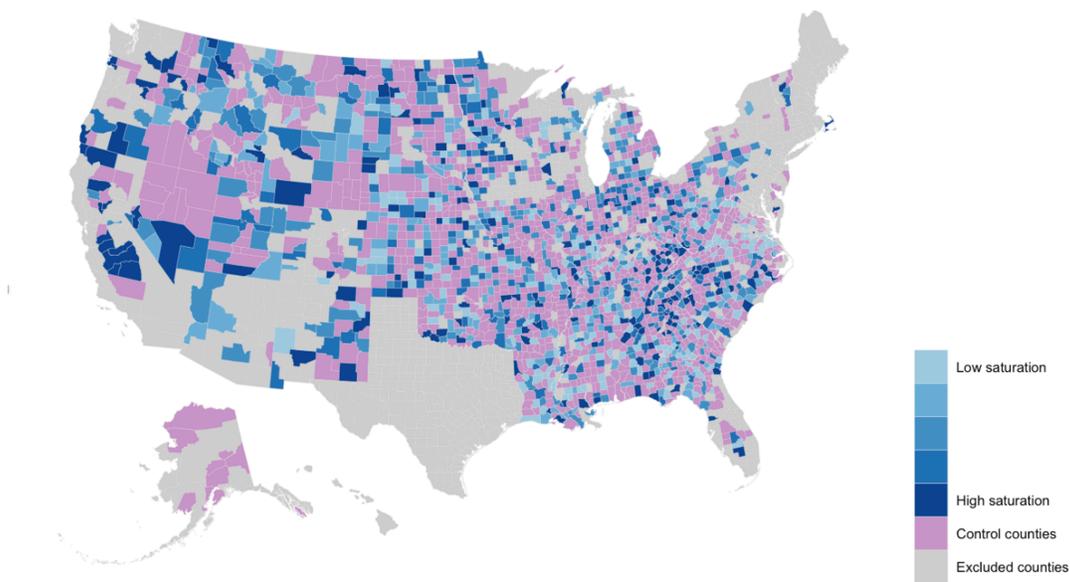

Figure 4: Geographic Distribution of Advertising Campaign

Notes: Gray regions are counties that were excluded (highly vaccinated, large population, or poor CDC records). Purple shows control counties. Blue shows the distribution of ads at the county level for treatment counties, with the five shades of blue corresponding to five quintiles of ad displays per resident.

# Supplemental Materials for "Using Donald Trump to Increase COVID-19 Vaccinations: Evidence from a Large-Scale Ad Experiment on YouTube" by Larsen, Ryan, Greene, Hetherington, Maxwell, and Tadelis, 2022

## A. Additional Design Details

### A.1 PSA Design

Our choice of wording for the final frame of the PSA, "Your vaccine is waiting for you," was driven by evidence that, for flu vaccines, patients are most responsive to framing suggesting that a vaccine is reserved for them (*26*).

### A.2. Selection of Treatment and Control Counties

We designed our experiment to administer ads such that they would be concentrated in areas with low vaccination rates while also facilitating our ability to estimate effects on actual vaccine uptake at the county level, the smallest geographic unit for which vaccine administrative records are widely available. Specifically, we excluded counties with vaccination rates above 50% (full vaccine series complete, according to CDC records as of September 28, 2021). We also excluded counties with populations above 1 million (45 counties). Aside from being culturally distinctive (e.g., large cities) and expensive to target, these had the potential to exert disproportionate influence on our results because the distribution of county populations has a long right tail. We also excluded the following other areas that we deemed inappropriate for our study. First, we excluded several—mostly uninhabited—Alaska Census areas. These are not conducive to YouTube targeting, as they are not counties. We excluded all counties in Texas (254), as these counties are not consistently included in CDC records (our source for measuring vaccine uptake). Third, we exclude Washington, D.C., given its unique cultural status.

These exclusions left us with 2,168 counties eligible for the study. We divided these into quintiles according to 1) population and 2) percentage of the population vaccinated and then created strata defined by the intersections of these two classifications. The creation of these strata happened before we chose to exclude low-vaccine counties from our experiment. Applying this restriction dropped the top quintile in terms of vaccination rates, leaving us with 20 strata created by the intersection of the two classifications. We then randomly assigned counties to treatment or control, blocking by the 20 strata, using the software developed by (*27*). This procedure resulted in 1,083 counties assigned to receive ads, and 1,085 retained as a control group. Table A.1 provides summary statistics at the county level for our full sample and separately for treatment and control counties.

### A.3. Ad Spending

We set our initial target price per 1,000 impressions (CPM) to $7.43, the level at which YouTube advertising algorithms and staff recommended. Between October 14th and 17th, we spent approximately $7,500 per day on ads. As the campaign progressed, and the pool of users to which Google's algorithm sent our ads changed dynamically, this initial target price proved too low to place our ad and spending dropped to between $2,000 and $3,000 on the 18th and 19th. Increasing our CPM target price to $10 allowed spending to surge above $8,000 per day from the 20th to the 24th. For the complete duration of the campaign, our average CPM was $8.55.

The total budget we spent on ads was $99,009.51. We spent the remainder of our original $100,000 budget in early testing to learn the platform's features. First, we spent $50 on October 8-9 on a small campaign that spread a small number of ad impressions across the whole country (not targeting any particular county). On October 11-12 we experimented with the county-targeting tools, spending $746 on ads. Some of the counties in our main campaign received ads at this point, but very few. The median number of impressions received by any county in our main study *before* October 14 was only 27, and there is no county in our main study in which more than 1% of the population could have seen the ad from our pre-October-14 experimentation.

Appendix Tables A.2 and A.3 contain additional summary statistics about the outlets through which ads were shown and characteristics of the users viewing and engaging with the ads. Each of these tables are constructed using data about the campaign from Google Ads tools.

### B. Estimation Details
### B.1 Intent-to-Treat Effect Estimation

Let $y_{it}$ denote the cumulative number of COVID-19 vaccine first doses among residents ages 12 and older in county $i$ up through date $t$, obtained from https://data.cdc.gov/Vaccinations/COVID-19-Vaccinations-in-the-United-States-County/8xkx-amqh. Our analysis encompasses dates from one month before the campaign to one month after, which includes 151,945 county-date observations. Let $Treat_i$ be an indicator variable equal to 1 if county $i$ is a treated county and 0 otherwise. Let $Post_t$ be a binary variable equal to 1 if date $t$ occurs on or after October 14, the start date of the campaign. Let $Population_i$ be the population of county $i$ (in units of 10,000). We estimate the following difference-in-difference regression:

$$y_{it} = \alpha + \lambda_t + \gamma_i + \beta(Treat_i \times Post_t) + \eta(Population_i \times Post_t) + \varepsilon_{it} \qquad (1)$$

The variable $\lambda_t$ is an effect for date $t$, capturing nationwide trends in vaccinations on a given date, and $\gamma_i$ is a county effect, capturing time-invariant differences in vaccination counts across counties. The interaction term $Population_i \times Post_t$ allows for the possibility that the cumulative vaccine count grows at a different rate over time in counties of different sizes. $\lambda_t$ and $\gamma_i$ also absorb the main effects $Treat_i$, $Population_i$, and $Post_t$. Our estimation algorithm to incorporate these two-way fixed effects relies on *(28)*. The residual $\varepsilon_{it}$ includes all unobserved factors affecting the number of vaccines administered in a particular county on a given date. As described in the body of the paper, not all counties in the treatment group received ads, and for those that did receive ads the exposure varied widely across counties. Thus, the primary coefficient of interest, $\beta$, is what is referred to as the *intent-to-treat* (ITT) effect.

One important assumption underlying our analysis is the Stable Unit Treatment Value Assumption (SUTVA), which requires that a county's treatment status does not affect the potential outcomes of other counties. This could be violated, for example, if some people view the ad in a treatment county and then cross county borders into a control county to get the vaccine. This particular type of violation would lead to our estimates of the intent-to-treat effect being understated.

The results of estimating equation (1) are shown in column 1 of Table 2. In column 2 we modify equation (1) to include flexible interactions of county population with dates in our sample period. These are implemented by replacing $\eta(Population_i \times Post_t)$ in equation (2) with

$$\sum_{\tau \in T} \eta_\tau \mathbf{1}_{t=\tau} + \eta_{\underline{t}} \mathbf{1}_{t<\underline{t}} + \eta_{\bar{t}} \mathbf{1}_{t>\bar{t}},$$

where $T = \{\underline{t}+1, \underline{t}+2, ..., t^*-1, t^*+1, ..., \bar{t}-2, \bar{t}-1\}$ is a window of dates defined as 13 days before the campaign up through 13 days after, with $t^*$, the date immediately preceding the campaign, being omitted. Thus, $\underline{t}$ is September 30, $\bar{t}$ is November 14, and $t^*$ is October 13. $\mathbf{1}_E$ is an indicator equal to 1 if the event $E$ is true. We also replicate these results in Table A.4, where we also display the estimated coefficient on the $Population_i \times Post_t$ interaction, which was omitted from the display in Table 2 for brevity.

**B.2 Average Causal Response (IV) Estimation**

As highlighted in Section 5 of the body of the paper, the actual ad exposure for treated counties varied widely, with some counties receiving little or no ads. Equation (1) measures the intent-to-treat effect; it does not allow us to measure the effect of *treatment intensity*. To measure the impact of increased ad exposure on vaccine uptake, it is tempting to replace the treatment dummy in equation (1), $Treat_i$, with the number of ads received by the county, which we denote $Ads_i$ (measured in units of 1,000). As Angrist and Imbens show, however, such an approach would be flawed (*20*). In our context this is because the number of ads is not necessarily exogenous, but determined by Google's algorithms. Angrist and Imbens propose a simple solution using *instrumental variables* (IV), instrumenting the exposure level $Ads_i$ with the treatment assignment dummy $Treat_i$. In our difference-in-difference regression framework, this becomes the following regression:

$$y_{it} = \alpha + \lambda_t + \gamma_i + \delta(Ads_i \times Post_t) + \eta(Population_i \times Post_t) + \varepsilon_{it} \qquad (2)$$

where we instrument for $Ads_i \times Post_t$ using $Treat_i \times Post_t$. The resulting estimate of $\delta$ is what Angrist and Imbens refer to as the *average causal response* (ACR). In this case, the estimate of $\delta$ represents the average causal response for treated counties from an additional 1,000 ad impressions. We rely on the estimation algorithm of *(29)*. Table 2, columns 3 and 4, report this estimate, with column 3 adding controls to column 1 for differential trends by population size, and column 4 doing the same to column 2. As in the case of the intent-to-treat effect, we also replicate these results in Table A.4, where we also display the estimated coefficient on the $Population_i \times Post_t$ interaction, which was omitted from the display in Table 2 for brevity.

In Table A.5, we replicate the IV analysis of regression (2) using the different treatment intensity metrics mentioned in Section 5.3. Panel A controls for $Population_i \times Post_t$ and panel B replaces this with the population and date dummy interactions, as in column 4 of Table 2. We normalize each variable by its standard deviation (across treatment counties) to facilitate comparison. Thus, each average causal response can be interpreted as the effect of a one-standard-deviation increase in the corresponding variable. In each specification in Table A.5, we find positive estimates that are statistically significant at the 0.05 level when standard errors are clustered at the state level and at the 0.20 level under county-level clustering. It is important to note that we do not analyze separately moving measures of treatment intensity; all of these measures move positively together (see correlations in Table A.6). The results in Table A.5 are therefore different ways of evaluating the average causal response using different measures of treatment intensity. Finally, note that in some Google Ads reports, in addition to ten-second-or-longer views, Google is unclear about whether *clicks* are also counted as *engagements* (also referred to in these tools as *interactions*). This has a negligible effect on our statistics for engagements, as clicks are rare in our data, occurring for less than 0.1 percent of ad impressions (see Table A.1).

**B.3 Robustness to Dropping Mis-recorded CDC Data**

During our sample period, two types of obvious mis-recording occur in the CDC county-level daily data. The first is that some counties show a decrease from one day to the next in their cumulative vaccination count. For example, the data can show that a given county has administered a total of 34,500 COVID-19 vaccine first doses since the beginning of the vaccine's availability, up through and including date $t$, and then show that this number decreases to 33,300 on the following day, which is impossible. The second type of error is simply that the vaccine count is missing for some dates during our sample for certain counties. We replicate our main analysis dropping any counties with mis-recorded CDC data. The results are shown in Table A.7. Columns 1-4 replicate the analysis of Table 2 but drop counties that have reported decreases in cumulative vaccine counts, and columns 5-8 instead drop counties that have missing vaccine counts for any date in the sample period. Table A.7 shows that the effects are positive, as in Table 2, but are slightly larger in magnitude. These results have a similar level of statistical significance to those in Table 2, although columns 5-8 are no longer significant at the 0.20 level under county-level clustering.

**B.4 Standard Errors**

As outlined in our pre-registration plan, we incorporate state-level clustering (43 clusters) to allow for the possibility that county-level residuals may be correlated within a state. The choice of the level at which to cluster standard errors is the subject of ongoing research in econometrics. It is known that bias can result from having too few clusters on one hand or from not clustering at an aggregate enough level on the other. However, the question of what constitutes "too few" is unresolved; some researchers point to 50 as a reasonable threshold, others to 20 *(30)*. According to *(30)*,

> The consensus is to be conservative and avoid bias and use bigger and more aggregate clusters when possible, up to and including the point at which there is concern about having too few clusters. For example, suppose your dataset included individuals within counties within states, and you were considering whether to cluster at the county level or the state level. We have been inclined to recommend clustering at the state level. If there was within-state cross-county correlation of the regressors and errors, then ignoring this correlation (for example, by clustering at the county level) would lead to incorrect inference.

Our choice of state-level clustering is driven by this conventional wisdom. More aggregate clustering typically results in larger standard errors, but, as highlighted in *(30)*, it is possible for more aggregate clustering to reduce standard errors when residuals are negatively correlated across observations in a more aggregate cluster. We find this to be the case in our setting, where the standard errors on our effects of interest are smaller under state-level clustering than county-level clustering.

Table A.8 shows our main results from Table 2 under our pre-registered specification for standard errors as well as under alternative specifications. Under county-level clustering, 95% confidence intervals contain zero, and results are significant only at the 0.20 level. When we allow for no clustering (and instead apply heteroskedasticity-robust standard errors) or clustering at the state-level, we reject the null hypothesis at the 0.05 level. We also reject the null hypothesis at the 0.10 level in columns 3-4 (but not in columns 1-2) when clustering at the level of a stratum (i.e., the 20 strata of counties from our stratified randomized sampling design). We also explore two-way clustering, following *(30)* and other work by these authors, combining geographic clustering with date-level clustering to allow for possible correlations across counties on a given date; Table A.8 demonstrates that this makes little or no difference to our estimated standard errors. In column 3, under stratum- and date-level clustering, the null is again rejected at the 0.05 level.

While our approach follows convention, we remark here that *(31)* critique this conventional view, presenting arguments for clustering at the level at which randomization occurs (the county, in our case). The authors do not consider our precise setting, however, accentuating that, in a setting with a time-series element (as is our case), the correct level at which to cluster is an open question.

An alternative approach to assessing statistical significance is randomization inference *(32)*, wherein the researcher uses monte carlo methods to simulate the distribution of effect sizes that arise under the "sharp null" hypothesis (i.e. a treatment effect of zero for all units). Under randomization inference for our ITT effect corresponding to column 1 of Table 2, we find a p-value of 0.17 from a two-tailed test. When we drop counties with negative vaccine count changes (as in column 1 of Table A.7), this p-value decreases to 0.15. Under a one-tailed test, randomization inference yields instead a p-value of 0.067, regardless of whether we exclude these counties with misrecorded data. These randomization inference p-values suggest that the state-level clustered standard errors are too small and that county-level standard errors are slightly too large. It bears emphasis that the assumptions underlying randomization inference differ from those of asymptotic approaches (our clustering approaches): where the asymptotic approaches test the null hypothesis that the *average* treatment effect across units is zero, randomization inference tests the more restrictive hypothesis that the effect is zero *for every unit*. Throughout the paper, we report both our pre-registered approach as well as county-level clustering to allow readers to consider confidence in rejecting the null hypothesis under alternative assumptions.

**B.5 Departures from Pre-Registration Plan**

We pre-registered our analysis plan via the Open Science Framework at https://osf.io/m9yhn/?view_only=c0d43e87224649e88b671eafddb22df8. Our analysis described in the body of the paper follows this pre-registered plan to the extent possible. Specifically, our pre-registered dependent variable is the number of vaccines administered in each county up through a given date. Our pre-registration plan also stated that we would analyze effects of our campaign through difference-in-difference OLS regressions, and we follow this plan throughout. The plan explained that we would estimate treatment effects on a sample of dates ranging from 14 days before to five days after the campaign, which we refer to here as our *restricted sample*. We also pre-registered an intention to explore wider date ranges, given uncertainty about how quickly treatment effects would onset.

After the campaign ran, however, we learned two unanticipated features of the data and of Google Ads algorithms that led to necessary modifications of our analysis. First, during the period of our study, county-level growth rates in vaccination counts were very different in counties of different population sizes, leading us to include controls for differential growth rates in counties of different sizes. And second, Google's algorithms distributed our ad very differently across target counties, leading us to incorporate an additional analysis (our IV approach) not described in our pre-registration plan. We discuss each of these points below.

We discovered, after our campaign, that in the time frame of our campaign counties of different population sizes grew in their vaccination counts at very different rates. To see this, we first estimate a version of equation (1) without including the $Population_i \times Post_t$ term. The results are shown in column 1 of Table A.9, where we observe a small point estimate (9.8) that is very imprecisely measured. The 95% confidence interval contains our preferred estimate from the body of the paper, 102.6 (from column 1 of Table 2). We then estimate a version of equation (1) without the $Treat_i \times Post_t$ interaction but including the $Population_i \times Post_t$ term. As in equation (1), the main effect of $Population_i$ is absorbed by county-level effects $\gamma_i$. The results are shown column 2 of Table A.9. We find a statistically significant and positive coefficient on the $Population_i \times Post_t$ term, implying that a county with 10,000 more residents has 275 more vaccinations in the post period. This increase is entirely independent of our experiment, as the results hold across all counties (column 2) and even within control counties alone (column 3). This suggests that, by not controlling for differential growth in vaccines in counties of different sizes, the specification in column 1 leaves a significant amount of statistical noise uncontrolled for. When we include the $Population_i \times Post_t$ interaction, we obtain the effect of 102.6 reported in column 1 of Table 2.

As outlined in our pre-registration plan, we estimated treatment effect first on a restricted sample of dates and then moved to wider date ranges. This restricted sample is a window from 14 days before to 5 days after the campaign, totaling to 72,815 county-date observations. Estimates of the intent-to-treat effect on this restricted sample are shown in columns 4-5 of Table A.9, with column 4 omitting the $Population_i \times Post_t$ interaction and column 5 including it. The results are too imprecisely measured to detect a significant effect the 0.05 level in either column, but in the latter the 95% confidence interval contains our preferred estimate from column 1 of Table 2. The final sample we chose to focus on is a wider date range, including dates from one month before the campaign to one month after, which we refer to as our *full sample* (the 151,945 county-date observations). We arrived at this window after exploring an *event study* version of equation (1) that allowed us to see the timing of the effects of the campaign. This event study is described in detail in Appendix C, and clearly reveals that the ad campaign affected vaccine counts with a lag, underlining the importance of allowing for a wider date range. Appendix C discusses several possible sources for this lag.

Our pre-registration plan did not describe the IV analysis we incorporate in Section 5 in the body of the paper. This is because, as with the large role played by county population or the lag in the effect of the campaign, we were not aware before the fact how drastically the ad exposure would vary across counties due to the nature of Google's ad auction platform. This variation in treatment intensity opened an additional informative lens of analysis for our study.

As a final, minor departure from our pre-registration plan, the plan described omitting the campaign start date (October 14) from our analysis. We replicated our analysis with and without this date and found that the qualitative and quantitative findings of the study were unchanged.

Though our primary results deviate from our pre-registered plan, we believe that our revised approach in analyses are all appropriate responses to changes in the research environment that could not be sufficiently anticipated at the time of the pre-registration. We believe, nonetheless, that they are in keeping with the spirit of our pre-registration plan and, under the circumstances, provide the most appropriate tools to assess the causal impact of our advertising campaign on vaccine uptake in the targeted counties.

## C. Event Study Analysis

We now describe an event study design that expands on our main regression analysis from the body of the paper to examine the intent-to-treat effect at each specific date before and after the campaign. We run the following regression:

$$y_{it} = \alpha + \lambda_t + \gamma_i + Treat_i\left(\beta_{\underline{t}}\mathbf{1}_{t\leq \underline{t}} + \sum_{\tau>\underline{t}, \tau\neq t^*}\beta_t \mathbf{1}_{t=\tau}\right) + Population_i\left(\eta_{\underline{t}}\mathbf{1}_{t\leq \underline{t}} + \sum_{\tau>\underline{t},\tau\neq t^*}\eta_t \mathbf{1}_{t=\tau}\right) + \varepsilon_{it} \quad (3)$$

Where the date $\underline{t}$ is September 30 and $t^*$ is October 13, the day before the start of the campaign. Thus, this specification estimates a date-specific effect for each date after September 30, and a single coefficient pooling together dates before that. The notation $\mathbf{1}_E$ is defined in Section B.

The results are shown in Figure A.1.A. Pointwise 95% confidence intervals (constructed using state-level clustering) are shown with the shaded region. The results suggest that the difference between treatment and control counties is not significantly different from zero before the campaign, which offers additional reassurance that our randomization worked as intended. This is a test of the *parallel trends* assumption required for identification in a difference-in-differences model, evidence that the number of vaccines in treatment counties – and the trend in that number – is not statistically significantly different from that of treatment counties prior to the start of the campaign. Figure A.1.A also shows that, after the campaign begins, the effect remains small initially, with confidence intervals containing zero. The cumulative effect increases near the end of the campaign and continues to increase slightly through the first two weeks following, peaking at around 100 vaccines—consistent with our main estimates from Table 2 in the body of the paper—at which point the cumulative effect decreases. Near the end of the sample, the difference between treatment and control counties is swamped by other noise, with the confidence intervals being quite wide and containing zero.

In Figure A.2.A, we replicate this analysis but with standard errors clustered at both the state and date level. While we find that this adjustment makes little difference our main standard error estimates (see Section B.4), in our event study, clustering at the date level (in addition to state) improves precision, allowing for contemporaneous correlations across counties. Other panels in Figure A.2 also parallel those in Figure A.1 using two-way clustering at the state and date level rather than only the state level. Figures A.3 and A.4 parallel Figures A.1 and A.2 but replacing state-level clustering with county-level clustering. As with the other results in the paper, the county-level clustering results in wider confidence bands than state-level clustering. Thus, the point estimates in each of Figures A.1 to A.4 are equivalent; only the confidence intervals differ.

There are several possible explanations for pattern of the point estimates (the solid black lines) we observe in Figure A.1.A. First, research on the effects of advertising on behavior suggests that there are important cumulative effects of exposure in that initially, several impressions may be needed to generate a response, after which the marginal impact of exposure declines (*33, 34*). Second, we discovered that the CDC data itself is recorded with a lag for some counties, implying that, if the treatment did have an effect on a given date, it may appear in CDC data on a later date. For example, in some observations in the CDC data, a county's cumulative vaccine count jumps up by over 30,000 in a single day—for one such county, this jump corresponds to about 1/3 of the entire county population—an implausibly large amount for a single day, but consistent with some counties updating their vaccine counts infrequently and, therefore, in batches.

Third, it is possible that those affected by the ad chose to *schedule* vaccine appointments for several days or a week in the future rather than getting vaccinated on the day they viewed the ad. Even for vaccines administered as walk-ins, rather than scheduled, appointments, it is plausible that the advertising campaign had an effect with a lag, with viewer's changing attitudes being reflected in actual vaccinations only after several days or a week of the attitude change. For example, viewers may decide, consciously or not, "I won't go out of my way to get the vaccine today, but I've decided to go ahead with getting vaccinated next time I find myself in Walgreens" (or CVS, Walmart, Kroeger, or Rite Aid—major distributors during this time period).

A fourth possibility is that the changes in our campaign itself on October 25 had an impact in its effectiveness. As explained in Section 3, on October 26 (in the afternoon), we excluded 18-24-year-olds from our targeting. In addition to directly shifting the age demographic, this may also have led to a shift in where our ad was placed; both of these shifts may have allowed our ad to reach more receptive viewers.

Finally, it may be that nation-wide changes in vaccine policy, which affected both treatment and control counties, had differential impacts on treatment counties due to the priming of the advertising campaign. The key change in vaccine policy that occurred during our sample period was that children ages 5-11 became eligible for the vaccine on Nov 2. Information about this policy change was leaked slowly over the weak preceding the change, with an official announcement from the CDC released on November 2 announcing the November 4 eligibility date. One possibility is therefore that the changing guidelines around childhood vaccination, and attention to these changes, can account for some of the reason that the treatment effect was larger near the end of the campaign.

It is not possible to completely rule out any of these possible explanations for the suddenness of the improvement in the effect around October 27. Regardless of the main drivers, the evidence in Figure A.1.A suggests that the campaign itself had a significant, temporary impact in increasing vaccinations.

We also estimate a variant of the above event study that controls for $y_{it-1}$, the lagged vaccine count within county $i$, on the right-hand side, as follows:

$$y_{it} = \alpha + \delta y_{it-1} + \lambda_t + \gamma_i + Treat_i\left(\beta_{\underline{t}}\mathbf{1}_{t\leq \underline{t}} + \beta_{\bar{t}}\mathbf{1}_{t\geq \bar{t}\}} + \sum_{\tau>\underline{t},\tau\neq t^*}\beta_t\,\mathbf{1}_{t=\tau}\right) +$$
$$Population_i\left(\eta_{\underline{t}}\mathbf{1}_{t\leq \underline{t}} + \eta_{\bar{t}}\mathbf{1}_{t\geq \bar{t}\}} + \sum_{\tau>\underline{t},\tau\neq t^*}\eta_t\,\mathbf{1}_{t=\tau}\right) + \varepsilon_{it} \qquad (4)$$

Specification (4) also bins together dates falling within the last two weeks of the sample (those on or after November 14, which is denoted $\bar{t}$. This regression estimates the difference between treatment and control counties in terms of their *daily* vaccination count on a given date rather than the cumulative vaccination count.

The results are shown in Figure A.1.C. These daily level results are more noisily estimated than the cumulative results from Figure A.1.A. However, consistent with panel A, Figure A.1.C shows that differences between treatment and control counties prior to the start of the campaign were small and largely insignificant. We observe large point estimates later in the campaign (in the last week of October), and then a leveling off at about 10 vaccines per day in the final days of the campaign and the two weeks afterward.

Upon investigation, we learned that the large spike on October 29 is driven by counties with misrecorded CDC data, in which the county records a *decrease* in its cumulative vaccination count over time, as discussed in Section B.3 above. Figure A.1.D omits these counties, and the October 29 spike disappears. For completeness, we also estimate the main event study, equation (3), on this subsample and display the results in Figure A.1.B. Here we see a similar pattern to that in panel A but with tighter confidence intervals. The fact that our point estimates are similar regardless of whether these counties are included, and that the confidence intervals are tighter without these counties, suggests that the data errors are uncorrelated with the treatment assignment and only introduce statistical noise, not bias, into our estimates.

**D. Heterogeneous Treatment Effects**

We now explore how the treatment effect and the response to the number of ads varies with county-level characteristics, including the percent of voters in the county who voted for Trump in 2016, the percent who are college educated, and the percent white. We use vote share for Trump from the 2016 election, as county-level measures are more readily available for this year; vote shares for the 2020 election are highly correlated with those of 2016. For each of these three characteristics, we compute the median across counties and we let $W_i$ be a dummy variable equal to 1 if county $i$ is below the median value for that characteristic. We then estimate regressions of the following form:

$$y_{it} = \alpha + \lambda_t + \gamma_i + \beta(Treat_i \times Post_t) + \phi(Treat_i \times Post_t \times W_i) + \psi(W_i \times Post_t) + \eta(Population_i \times Post_t) + \varepsilon_{it} \quad (5)$$

In equation (5), $\beta$ represents the intent-to-treat effect for above-median counties and $\beta + \phi$ represents the effect for below-median counties. As in the body of the paper, we follow (*20*) to also estimate the average causal response using instrument variables. Specifically, we replace $Treat_i$ in equation (5) with $Ads_i$ and we instrument for $(Ads_i \times Post_t)$ and $(Ads_i \times Post_t \times W_i)$ using $(Treat_i \times Post_t \times W_i)$ and $(Treat_i \times Post_t)$.

The estimates from these regressions are shown in Table A.10. In this table, $W_i$ represents a county being below the median in terms of the Trump vote share in columns 1–2, the percent who are college educated in columns 3–4, and the percent white in columns 5–6. Column 1 shows that counties with a smaller Trump vote share are responsive to the treatment, with an ITT effect of 260 vaccines. This estimate is higher than the effect for the average county (our main results in Table 2) and is significant at the 0.01 level under state-level clustering and at the 0.10 level under county-level clustering. The estimated average causal response to 1,000 ads is 17.53 vaccines, also larger in magnitude than our estimate from Table 2. This estimate is significant at the 0.01 level under state-level clustering and at the 0.20 level under county-level clustering. For counties with an above-median Trump share, there is no significant response to the treatment—the $Treat \times Post$ effect is insignificant. It is important to emphasize that, because all the counties in our treatment and control samples have low vaccination rates, and because Republicans have demonstrated greater vaccine hesitancy than Democrats on average, Trump-favoring counties are over-represented in our sample, and thus the median county in our sample had a Trump vote share of 70%. What these results suggest is that our message is highly effective in garnering a behavioral response among certain counties (those with less than 70% of voters favoring Trump) and not among those with extreme proportions of Trump supporters. In columns 3-6, we find no significant differences in the response of counties based on the education level or racial makeup.

### E. Measuring Vaccines and Ads in Rates

Here we evaluate an alternative specification for regressions (1) and (2) where we replace the dependent variable with the *percent* of residents in county $i$ who are vaccinated up through date $t$. In equation (2) we also replace the number of ads with the number of ads per 100 county residents. These specifications allow us to control for county population by normalizing (dividing by county population) rather than controlling for it additively as in regressions (1) and (2) in the body of the paper.

The results are displayed in Table A.11. Columns 1-2 show the ITT effect and columns 3-4 the ACR. For comparison, even columns control for the $(Population_i \times Post_t)$ interaction and odd columns omit this interaction. The number of observations is slightly higher in Table A.11 than in our main analysis (163,856 county-date observations rather than 151,945) because, for some observations, the vaccination count is missing on certain dates in the CDC data even though the vaccination rate is recorded.

All effects are more imprecisely measured than are our main results, which are evaluated in levels (Table 2). Indeed, although the point estimates themselves are positive, they are only significant at the 0.20 level, regardless of the level of clustering. If these numbers were significant, the column 1 estimates would suggest the campaign increased the percent vaccinated (a variable ranging from 0 to 100) by 0.57 percentage points in the average county. Similarly, the column 3 results would suggest that an increase of 1 more ad per 100 county residents increases the percent vaccinated by 0.03 percentage points. But, again, these estimates are not significant at conventional levels. Comparing odd and even columns, we see very similar point estimates, suggesting that controlling directly for county population after normalizing by county population does little to affect the estimates.

As demonstrated in the histograms in Figures 3.A and 3.B and in the surrounding discussion, the number of ad impressions varies widely across counties, but the number of ads per capita does as well, and thus the decision of whether to measure ads in rates or levels has no single correct answer, as the actual meaning of being "treated" is not fully captured by either. As discussed in Sections 5.3 and B.2, we find a positive and statistically significant effect (under state-level clustering) of treatment intensity on vaccines whether measured as the number of ads or the number of ads per 100 residents (see Table A.5). The results of Table A.11 are therefore illustrative primarily of the difference in the dependent variable: the vaccine count vs. vaccines per capita. Our preferred specification, which we pre-registered, measures vaccines in levels rather than in rates, because our primary research question is whether our ad campaign changed behavior sufficiently to increase the total count of vaccinated individuals (regardless of how large this change is relative to county size, which is what the effect in rates measures).

**F. Survey Instrumentation**

We contracted with Qualtrics to gather six different 2,400 respondent samples at regular intervals between April 2020 and September 2021 to measure public opinion about COVID-19. Qualtrics conducts Census-targeted sampling to ensure that survey samples closely match the U.S. benchmarks for age, education, gender, and race/ethnicity. The instrument fielded in March/April, as well as August/September, each asked respondents how much confidence they had in several sources "when it comes to advising you on taking the COVID-19 vaccine." In the March/April wave, 422 participants categorized themselves as Republicans and as unvaccinated (i.e., they did not select "I have already been vaccinated" in response to the question below). In the August/September wave, this number was 387. The exact wording for questions used in the Table 1 analysis appears below.

*How much confidence do you have in each of the following when it comes to advising you on taking the COVID-19 vaccine?*

|  | A great deal of confidence (1) | A fair amount (2) | Not too much (3) | No confidence at all (4) | Not Applicable (5) |
|---|:---:|:---:|:---:|:---:|:---:|
| Donald Trump | o | o | o | o | o |
| Joe Biden | o | o | o | o | o |
| Director of the U.S. National Institute of Allergy and Infectious Diseases Dr. Anthony Fauci | o | o | o | o | o |
| Your personal doctor | o | o | o | o | o |
| The scientific community | o | o | o | o | o |

[March/April 2021 version] *You may have heard that a COVID-19 vaccine has been authorized for distribution in the United States. If the vaccine were available for you to take today, would you*

- o Definitely get the vaccine
- o Probably get the vaccine
- o Probably NOT get the vaccine
- o Definitely NOT get the vaccine
- o I have already been vaccinated

[August/September 2021 version] *You likely heard that a COVID 19 vaccine has been authorized for distribution in the United States. How likely are you to get one of the COVID 19 vaccines?*

- o Definitely will get the vaccine
- o Probably will get the vaccine
- o Probably will NOT get the vaccine
- o Definitely will NOT get the vaccine
- o I have already received at least one dose

*Generally speaking, do you usually think of yourself as a DEMOCRAT, a REPUBLICAN, an INDEPENDENT, or what?*

- o Democrat
- o Republican
- o Independent
- o No preference
- o Other party (specify) \_\_\_\_\_\_\_\_\_\_\_\_\_\_\_\_\_\_\_\_\_\_\_\_\_\_\_\_\_\_\_\_\_\_\_\_\_\_\_\_

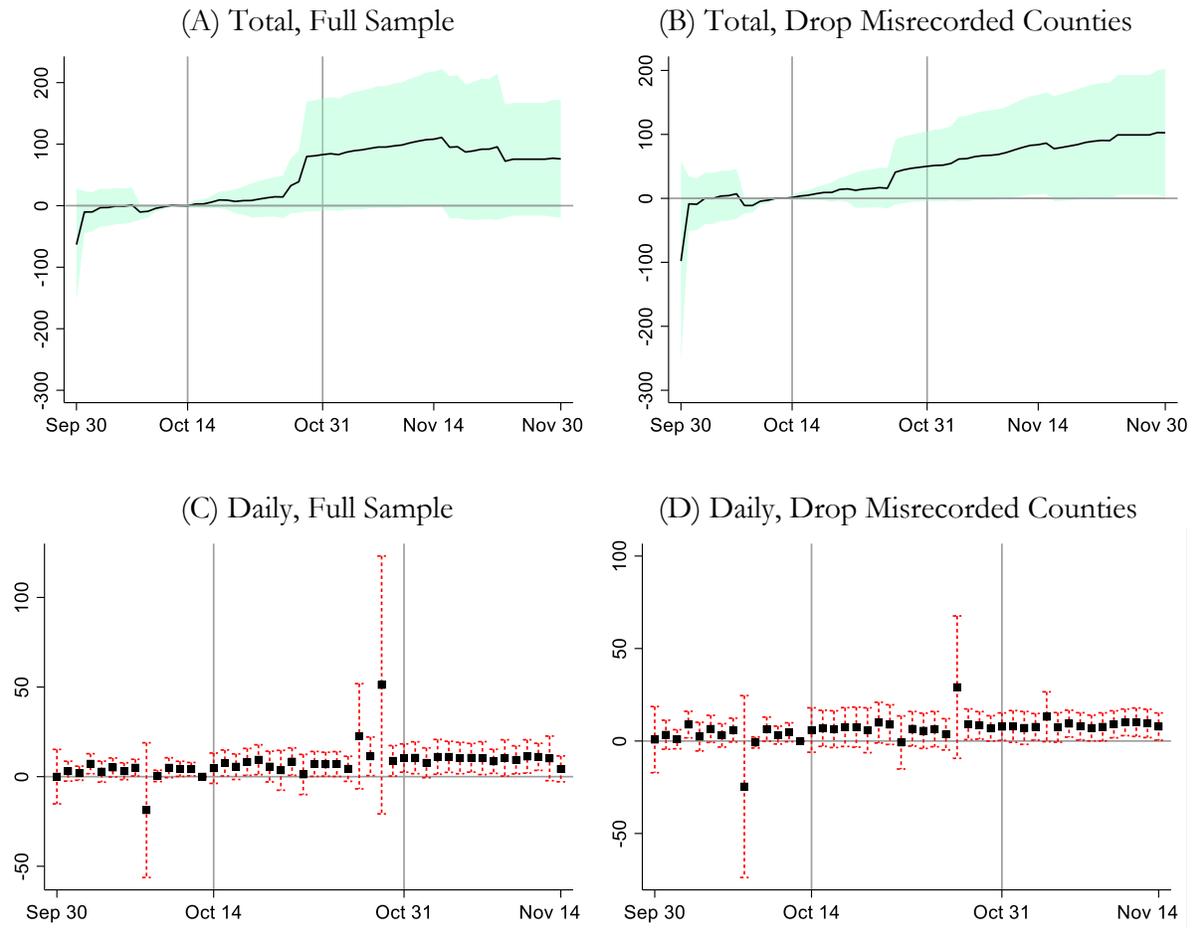

Figure A.1: Event Study with Standard Errors Clustered at State Level

Notes: Panels A and B display coefficients from event study regression (3), the effect on the cumulative vaccine count up through a given date. Panels C and D display results from regression (4), the effect on the number of vaccines administered on a given date. Panels on the left use the full sample and those on the right drop counties that ever record a decrease in cumulative vaccine count over time. Shaded region in panels A and B, and red dashed lines in panels C and D, representing pointwise 95% confidence intervals computed under clustering at the state level.

Figure A.2: Event Study with Standard Errors Clustered at State and Date Levels

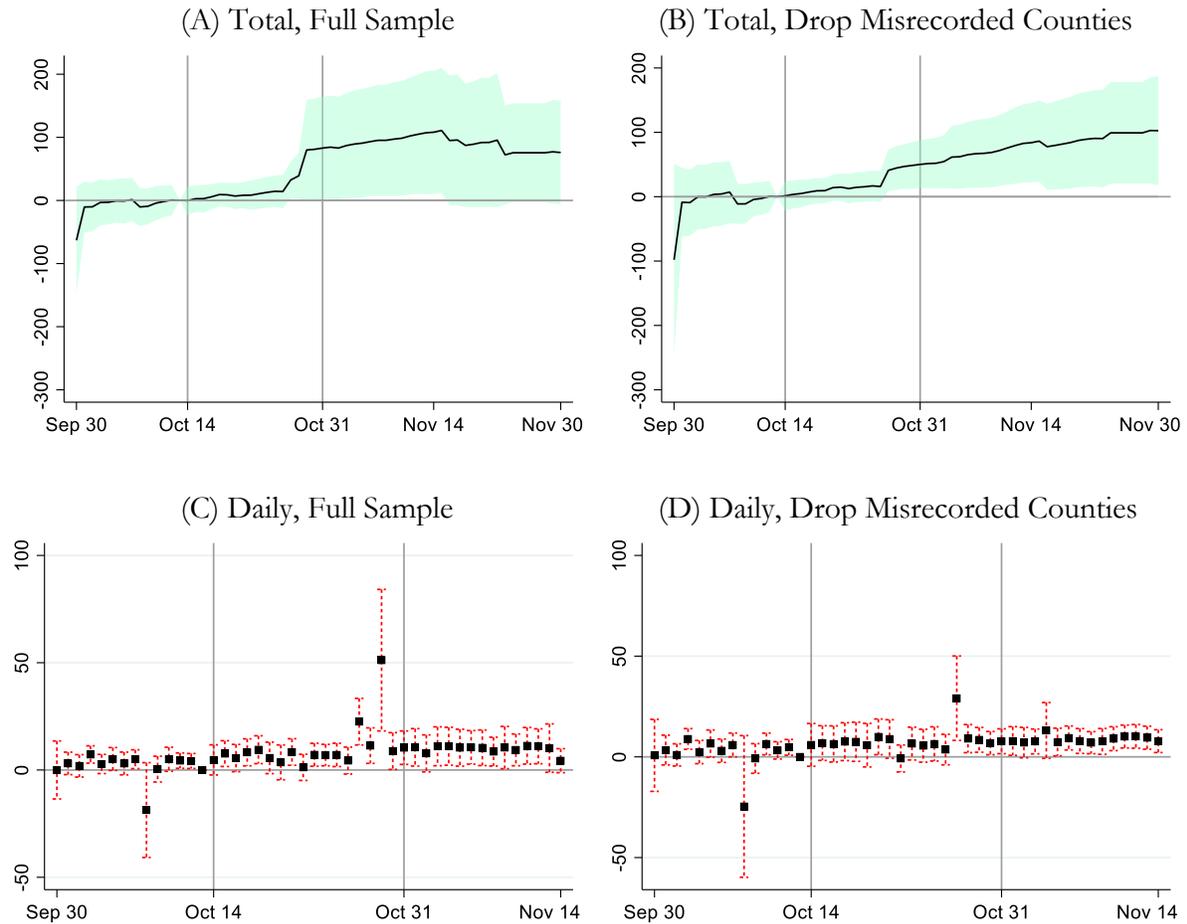

Notes: Panels A and B display coefficients from event study regression (3), the effect on the cumulative vaccine count up through a given date. Panels C and D display results from regression (4), the effect on the number of vaccines administered on a given date. Panels on the left use the full sample and those on the right drop counties that ever record a decrease in cumulative vaccine count over time. Shaded region in panels A and B, and red dashed lines in panels C and D, representing pointwise 95% confidence intervals computed under two-way clustering at the state level and date level.

Figure A.3: Event Study with Standard Errors Clustered at County Level

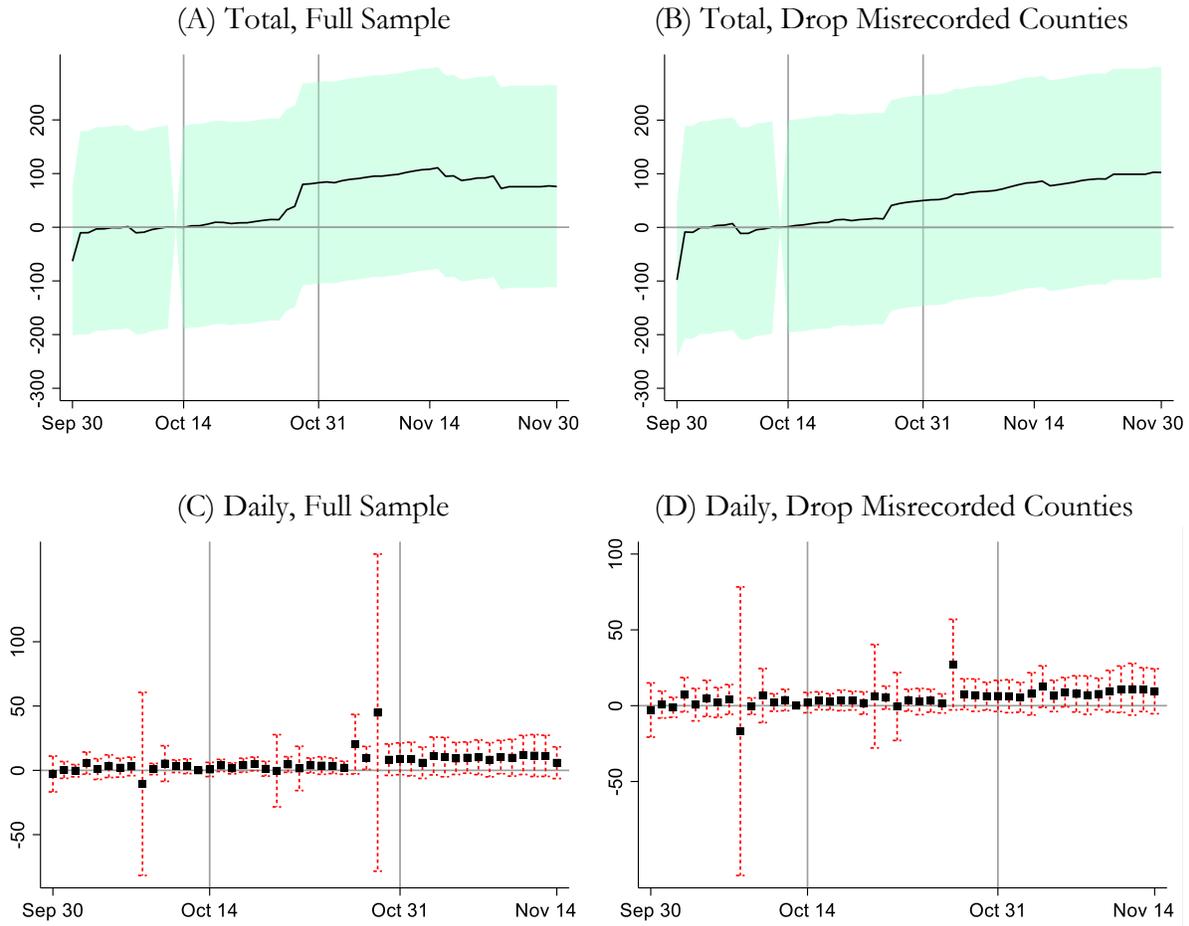

Notes: Panels A and B display coefficients from event study regression (3), the effect on the cumulative vaccine count up through a given date. Panels C and D display results from regression (4), the effect on the number of vaccines administered on a given date. Panels on the left use the full sample and those on the right drop counties that ever record a decrease in cumulative vaccine count over time. Shaded region in panels A and B, and red dashed lines in panels C and D, representing pointwise 95% confidence intervals computed under clustering at the county level.

Figure A.2: Event Study with Standard Errors Clustered at County and Date Levels

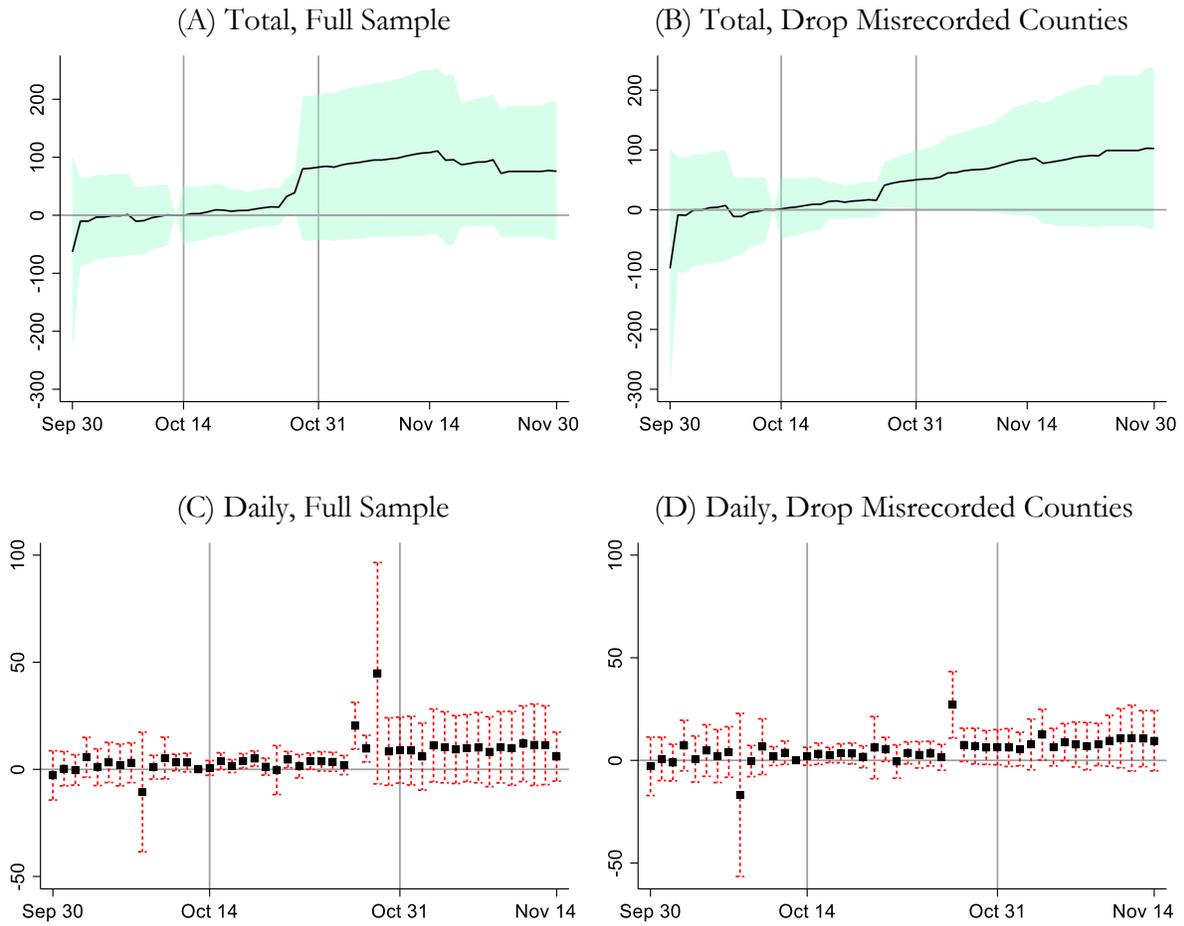

Notes: Panels A and B display coefficients from event study regression (3), the effect on the cumulative vaccine count up through a given date. Panels C and D display results from regression (4), the effect on the number of vaccines administered on a given date. Panels on the left use the full sample and those on the right drop counties that ever record a decrease in cumulative vaccine count over time. Shaded region in panels A and B, and red dashed lines in panels C and D, representing pointwise 95% confidence intervals computed under two-way clustering at the county level and date level.

Table A.1: Descriptive Statistics by County

|  | Mean | Std. Dev. | Min | Median | Max |
|---|---|---|---|---|---|
| **All Counties** | | | | | |
| County Population (10,000s) | 4.72 | 8.75 | .0463 | 2.16 | 99.9 |
| Vaccine First Doses Before Campaign | 24,123 | 48,655 | 92 | 10,027 | 591,758 |
| Number of Counties | 2,168 | | | | |
| **Treatment Counties** | | | | | |
| County Population (10,000s) | 4.56 | 8.12 | .0463 | 2.08 | 99.9 |
| Vaccine First Doses Before Campaign | 23,282 | 45,442 | 92 | 9,704 | 591,758 |
| Number of Ads | 10,679 | 25,245 | 0 | 3,764 | 346,089 |
| Number of Ads per 100 Residents | 19.3 | 9.17 | 0 | 19 | 160 |
| Interaction Rate | 41.3 | 3.32 | 0 | 41.1 | 63.3 |
| View Rate | 12.4 | 1.49 | 0 | 12.3 | 25 |
| Click Rate | .0838 | .0769 | 0 | .0776 | 1.14 |
| Cost (dollars) per 1,000 Ads (CPM) | 8.53 | .406 | 0 | 8.55 | 10 |
| Number of Counties | 1,083 | | | | |
| **Control Counties** | | | | | |
| County Population (10,000s) | 4.88 | 9.33 | .0465 | 2.24 | 96.5 |
| Vaccine First Doses Before Campaign | 24,954 | 51,643 | 216 | 10,435 | 530,314 |
| Number of Counties | 1,085 | | | | |

Notes: Descriptive statistics at the county level. Top panel shows statistics for all counties, middle panel for treatment counties, and bottom panel for control counties. Vaccine first doses before campaign correspond to the vaccine count in the county on October 13. A county's engagement rate is the number of 10-second (or longer) views of the ad, divided by the number of ad impressions the county received, multiplied by 100. View rate is the number of complete views of the ad divided by the number of impressions, multiplied by 100. Click rate is the number of times the Fox News story link was clicked, divided by the number of ads, multiplied by 100. CPM is the average cost (in dollars) of purchasing 1,000 ad impressions in the county.

Table A.2: Where the Ad Placed

|  | # Outlets | Total Ads | % View 10 Sec. | % View Full | Cost ($) per 1,000 Ads |
|---|---|---|---|---|---|
| A. Outlet Type | | | | | |
| Mobile application | 10,072 | 2,292,337 | 31.11 | 6.83 | 7.81 |
| Site | 840 | 94,992 | 37.00 | 13.71 | 6.07 |
| YouTube channel | 150,284 | 7,760,170 | 44.52 | 16.33 | 8.68 |
| B. Outlet Characteristics | | | | | |
| Contains "fox" | 235 | 270,293 | 49.85 | 17.81 | 8.48 |
| Contains "news" | 1,479 | 550,878 | 49.28 | 17.50 | 8.47 |
| C. Top YouTube Channels by Number of Ads Placed | | | | | |
| Fox News | | 214,920 | 50.73 | 18.32 | 8.41 |
| Forbes Breaking News | | 67,667 | 48.74 | 16.39 | 8.82 |
| YouTube Movies | | 34,934 | 51.67 | 19.37 | 9.02 |
| penguinz0 | | 34,158 | 37.22 | 10.49 | 8.39 |
| NFL | | 33,502 | 42.45 | 13.78 | 8.44 |
| Markiplier | | 24,844 | 43.15 | 13.34 | 8.75 |
| Inside Edition | | 23,947 | 36.70 | 8.92 | 8.96 |
| NewsNation Now | | 23,916 | 50.28 | 18.97 | 8.67 |
| Sky News Australia | | 23,218 | 52.14 | 17.05 | 8.46 |
| NBC News | | 22,030 | 52.41 | 22.11 | 8.30 |
| D. Type of Device of Viewer | | | | | |
| TV screens | | 3,510,664 | 46.00 | 16.84 | 8.15 |
| Tablets | | 1,555,365 | 39.19 | 10.72 | 8.34 |
| Computers | | 447,873 | 37.79 | 9.35 | 8.77 |
| Mobile phones | | 6,059,672 | 36.99 | 10.42 | 8.83 |

Notes: Descriptive statistics for the ad campaign from Google Ads tools. Panel A displays characteristics of the campaign for different types of outlets where the ad placed. Panel B displays characteristics for outlets with "fox" or "news" in the outlet name. Panel C displays characteristics of the top 10 YouTube channels (ranked by the number of impressions) on which the ad placed. Panel D displays the type of electronic device on which the user viewed the ad. The total number of ads in panel A sums to slightly less than 11.6 million because, for some ad impressions, Google Ads tools did not provide details on the outlet type.

Table A.3: Descriptive Statistics for YouTube Viewers

|  | Total Ads | % View 10 Sec. | % View Full | Cost ($) per 1,000 Ads |
|---|---|---|---|---|
| A. Gender | | | | |
| Female | 2,963,754 | 40.80 | 12.03 | 8.69 |
| Male | 5,290,982 | 43.76 | 12.78 | 8.96 |
| Unknown | 3,318,838 | 33.46 | 12.02 | 7.79 |
| B. Age | | | | |
| 18 - 24 | 1,593,423 | 38.94 | 11.09 | 8.69 |
| 25 - 34 | 1,189,531 | 41.16 | 12.32 | 9.25 |
| 35 - 44 | 1,010,042 | 43.15 | 13.26 | 9.16 |
| 45 - 54 | 1,197,026 | 43.28 | 13.05 | 8.83 |
| 55 - 64 | 1,615,427 | 43.64 | 12.67 | 8.60 |
| 65+ | 1,375,389 | 46.26 | 12.95 | 8.69 |
| Unknown | 3,592,736 | 34.23 | 12.11 | 7.93 |
| C. Income | | | | |
| Top 10% | 340,894 | 42.95 | 11.50 | 9.67 |
| 11–20% | 572,874 | 44.17 | 12.24 | 9.55 |
| 21–30% | 766,789 | 43.70 | 12.51 | 9.56 |
| 31–40% | 646,236 | 43.72 | 12.31 | 9.44 |
| 41–50% | 548,432 | 44.54 | 12.79 | 9.40 |
| Lower 50% | 2,720,921 | 45.24 | 12.75 | 9.18 |
| Unknown | 5,977,428 | 35.85 | 12.21 | 7.81 |
| D. Parental Status | | | | |
| Parent | 1,449,921 | 43.95 | 12.71 | 9.55 |
| Not a parent | 3,987,477 | 44.26 | 12.17 | 9.34 |
| Unknown | 6,136,176 | 36.39 | 12.42 | 7.81 |

Notes: Descriptive statistics on ad viewers from Google Ads tools, including the total number of ad impressions, the percentage watching at least 10 seconds or the full ad, and the cost per 1,000 ad impressions (CPM). Panel A displays statistics by gender, panel B by age, panel C by income level, and panel D by parental status. In each panel, "unknown" indicates users for which Google does not know a given characteristic.

Table A.4: Vaccine Increase Per County: Unreported Table 2 Estimates

|  | Intent-to-Treat Effect | | Average Causal Response of 1,000 Ads | |
|---|---|---|---|---|
|  | (1) | (2) | (3) | (4) |
| Treat × Post | 102.6 | 101.4 |  |  |
|  | (46.65)** | (46.72)** |  |  |
|  | (78.74)+ | (78.76)+ |  |  |
|  |  |  |  |  |
| Ads × Post |  |  | 8.606 | 8.500 |
|  |  |  | (3.986)** | (3.992)** |
|  |  |  | (6.608)+ | (6.609)+ |
|  |  |  |  |  |
| Pop. × Post | 275.4 |  | 264.5 |  |
|  | (32.67)*** |  | (32.41)*** |  |
|  | (18.71)*** |  | (21.68)*** |  |
| County Fixed Effects | Yes | Yes | Yes | Yes |
| Date Fixed Effects | Yes | Yes | Yes | Yes |
| Pop. × Post Dummy | Yes |  | Yes |  |
| Pop. × Date Dummies |  | Yes |  | Yes |
|  |  |  |  |  |
| Observations | 151945 | 151945 | 151945 | 151945 |

Notes: Regression results corresponding to Table 2, but with $Population_i \times Post_t$ interaction also displayed. All regressions include fixed effects at the county and date levels. Columns 1-2 correspond to regression (1) and columns 3-4 correspond to regression (2). Columns 2 and 4 replaces the $Population_i \times Post_t$ interaction with interactions of county population with (i) dummies for each date within two weeks before to two weeks after the campaign (omitting the date before the campaign started), (ii) a dummy variable for two weeks or more before, and (iii) a dummy variable for two weeks or more after. "***", "**", "*", and "+" indicate significance (from a two-tailed test) at the 0.01, 0.05, 0.10, and 0.20 levels. Standard errors, reported in parentheses below each estimate, are clustered at the state level in the first row and at the county level in the second row.

Table A.5: Vaccine Increase Per County Under Alternative Measures of Treatment Intensity

| Treatment Intensity Measure | Engagement Rate (1) | View Rate (2) | Click Rate (3) | Ads per 100 Residents (4) | CPM (5) |
|---|---|---|---|---|---|
| **A.** | Controlling for Population × Post Dummy | | | | |
| Average Causal Response | 8.255 | 12.34 | 94.12 | 48.37 | 4.877 |
| | (3.745)** | (5.615)** | (43.14)** | (22.10)** | (2.217)** |
| | (6.333)+ | (9.467)+ | (72.11)+ | (37.10)+ | (3.742)+ |
| | | | | | |
| Pop. × Post | 275.5 | 275.4 | 275.0 | 274.8 | 275.4 |
| | (32.67)*** | (32.66)*** | (32.58)*** | (32.65)*** | (32.67)*** |
| | (18.70)*** | (18.70)*** | (18.74)*** | (18.77)*** | (18.71)*** |
| | | | | | |
| **B.** | Controlling for Population × Date Dummies Average | | | | |
| Causal Response | 8.153 | 12.19 | 92.96 | 47.77 | 4.817 |
| | (3.751)** | (5.623)** | (43.20)** | (22.13)** | (2.220)** |
| | (6.335)+ | (9.469)+ | (72.12)+ | (37.10)+ | (3.742)+ |
| County Fixed Effects | Yes | Yes | Yes | Yes | Yes |
| Date Fixed Effects | Yes | Yes | Yes | Yes | Yes |
| | | | | | |
| Observations | 151945 | 151945 | 151945 | 151945 | 151945 |

Notes: Regression results from IV regressions, as in columns 3-4 of Table 2, but using different treatment intensity measures instead of number of ads. The dependent variable is the same as in Table 2, but the measure of treatment intensity differs by column. Each variable is normalized by dividing by its standard deviation across counties. A county's engagement rate is the number of 10-second (or longer) views of the ad, divided by the number of ad impressions the county received, multiplied by 100. View rate is the number of complete views of the ad divided by the number of impressions, multiplied by 100. Click rate is the number of times the Fox News story link was clicked, divided by the number of ads, multiplied by 100. CPM is the average cost (in dollars) of purchasing 1,000 ad impressions in the county. Panel A controls for a $Population_i \times Post_t$ and panel B replaces the $Population_i \times Post_t$ interaction with interactions of county population with date dummies, as in column 4 of Table 2. "***", "**", "*", and "+" indicate significance (from a two-tailed test) at the 0.01, 0.05, 0.10, and 0.20 levels. Standard errors, reported in parentheses below each estimate, are clustered at the state level in the first row and at the county level in the second row.

Table A.6: Pairwise Correlations Between Measures of Treatment Intensity

| | # Ads | # Ads per 100 Residents | Engagement Rate | View Rate | Click Rate |
|---|---|---|---|---|---|
| # Ads per 100 Residents | 0.452 | | | | |
| Engagement Rate | 0.268 | 0.820 | | | |
| View Rate | 0.283 | 0.824 | 0.993 | | |
| Click Rate | 0.224 | 0.529 | 0.604 | 0.601 | |
| CPM | 0.288 | 0.829 | 0.996 | 0.988 | 0.613 |

Notes: Table displays pairwise correlations between each measure of treatment intensity used in Table A.5 as well as the primary measure of treatment intensity (the number of ads) used in Table 2.

Table A.7: Regression Results Omitting Counties with Mis-recorded Data

|  | Counties Without Negative Counts | | | | Counties Without Missing Counts | | | |
|---|---|---|---|---|---|---|---|---|
|  | ITT Effect | | ACR of 1,000 Ads | | ITT Effect | | ACR of 1,000 Ads | |
|  | (1) | (2) | (3) | (4) | (5) | (6) | (7) | (8) |
| Treat × Post | 114.2 | 112.9 | | | 103.3 | 103.3 | | |
|  | (51.79)** | (51.83)** | | | (48.94)** | (48.95)** | | |
|  | (83.44)+ | (83.46)+ | | | (80.89) | (80.91) | | |
| Ads × Post | | | 9.014 | 8.914 | | | 8.633 | 8.633 |
|  | | | (4.076)** | (4.079)** | | | (4.172)** | (4.173)** |
|  | | | (6.608)+ | (6.609)+ | | | (6.768) | (6.769) |
| Pop. × Post | 285.5 | | 273.8 | | 276.5 | | 265.6 | |
|  | (34.08)*** | | (33.39)*** | | (32.93)*** | | (32.48)*** | |
|  | (20.24)*** | | (23.17)*** | | (18.93)*** | | (21.94)*** | |
| County Effects | Yes | Yes | Yes | Yes | Yes | Yes | Yes | Yes |
| Date Effects | Yes | Yes | Yes | Yes | Yes | Yes | Yes | Yes |
| Pop. × Post | Yes | | Yes | | Yes | | Yes | |
| Pop. × Dates | | Yes | | Yes | | Yes | | Yes |
| Observations | 116163 | 116163 | 116163 | 116163 | 144144 | 144144 | 144144 | 144144 |

Notes: Columns 1-4 report regression results from the same specifications as in Table 2 but using only counties in which CDC records do not show a decrease in the cumulative vaccination count for any date. Columns 5-8 report results as in Table 2 but using only counties in which CDC records are not missing for any date. ITT refers to intent-to-treat and ACR to average causal response. "***", "**", "*", and "+" indicate significance (from a two-tailed test) at the 0.01, 0.05, 0.10, and 0.20 levels. Standard errors, reported in parentheses below each estimate, are clustered at the state level in the first row and at the county level in the second row.

Table A.8: Standard Errors for Alternative Clustering Levels

|  | Intent-to-Treat Effect | | Average Causal Response of 1,000 Ads | |
|---|---|---|---|---|
|  | (1) | (2) | (3) | (4) |
| Effect | 102.6 | 101.4 | 8.606 | 8.5 |
| Alternative Standard Errors | | | | |
| Heteroskedasticity-robust | (20.15)*** | (18.78)*** | (1.69)*** | (1.58)*** |
| County Clustering | (78.74)+ | (78.76)+ | (6.61)+ | (6.61)+ |
| State Clustering | (46.65)** | (46.72)** | (3.99)** | (3.99)** |
| Stratum Clustering | (69.76)+ | (69.75)+ | (4.51)* | (4.52)* |
| County and Date Clustering | (77.38)+ | (77.94)+ | (6.49)+ | (6.52)+ |
| State and Date Clustering | (47.61)** | (47.85)** | (3.95)** | (3.98)** |
| Stratum and Date Clustering | (69.21)+ | (68.83)+ | (4.35)** | (4.36)* |
| County Fixed Effects | Yes | Yes | Yes | Yes |
| Date Fixed Effects | Yes | Yes | Yes | Yes |
| Pop. × Post Dummy | Yes |  | Yes |  |
| Pop. × Date Dummies |  | Yes |  | Yes |

Notes: Regression results corresponding to Table 2 with alternative standard error estimates, including clustering at the county level (2,032 counties), state level (43 states), or stratum level (20 strata). Heteroskedasticity-robust standard errors use no clusters. The final three rows of standard errors use two-way clustering at the geographical level and at the data level. State clustering corresponds to the standard error estimates reported in Table 2. "***", "**", "*", and "+" indicate that, under the specified standard error estimates, the effect in a given column is significant (from a two-tailed test) at the 0.01, 0.05, 0.10, or 0.20 levels, respectively.

Table A.9: Intent-to-Treat Effect Under Alternative Specifications

|  | (1) | (2) | (3) | (4) | (5) |
|---|---|---|---|---|---|
| Treat × Post | 9.758 |  |  | -1.426 | 43.37 |
|  | (86.38) |  |  | (42.23) | (26.83)+ |
|  | (139.6) |  |  | (68.07) | (39.88) |
| Pop. × Post |  | 275.3 | 273.1 |  | 131.0 |
|  |  | (32.71)*** | (35.25)*** |  | (16.52)*** |
|  |  | (18.71)*** | (27.56)*** |  | (9.936)*** |
| County Effects | Yes | Yes | Yes | Yes | Yes |
| Date Effects | Yes | Yes | Yes | Yes | Yes |
| Sample | Full | Full | Control | Restricted | Restricted |
| Observations | 151945 | 151945 | 76266 | 72815 | 72815 |

Notes: Regression results. Columns 1-2 use the full sample, column 3 uses only control counties, and columns 4-5 use the restricted sample period (14 days before to five days after the campaign). All regressions include fixed effects at the county and date levels. Column 1 runs a version of regression (1) without the $Population_i \times Post_t$ interaction, whereas columns 2-3 run regression (1) without the $Treat_i \times Post_t$ interaction. Columns 4 repeats the specification of column 1, and column 5 repeats the specification of Table 2 column 1, but on the restricted sample of dates. "***", "**", "*", and "+" indicate significance (from a two-tailed test) at the 0.01, 0.05, 0.10, and 0.20 levels. Standard errors, reported in parentheses below each estimate, are clustered at the state level in the first row and at the county level in the second row.

Table A.10: Vaccine Increase Per County: Heterogeneous Effects and Causal Responses

|  | W = Below Median % Trump | | W = Below Median % College | | W = Below Median % White | |
| --- | --- | --- | --- | --- | --- | --- |
|  | (1) | (2) | (3) | (4) | (5) | (6) |
| Treat × Post × W | 260.0 (80.43)*** (154.4)* |  | -56.76 (152.2) (158.5) |  | 216.3 (172.9) (152.3)+ |  |
| Ads × Post × W |  | 17.53 (6.361)*** (11.19)+ |  | 4.496 (12.54) (10.84) |  | 12.15 (16.82) (12.18) |
| Treat × Post | -27.05 (29.54) (45.95) |  | 131.1 (112.7) (152.6) |  | -4.971 (65.86) (54.66) |  |
| Ads × Post |  | -4.307 (4.701) (7.557) |  | 7.168 (6.588) (8.499) |  | -0.377 (11.33) (9.545) |
| W × Post | 69.90 (201.2) (95.13) | 140.4 (192.5) (76.04)* | -74.95 (236.9) (104.6) | -126.7 (225.4) (80.92)+ | 144.7 (181.6) (95.52)+ | 212.4 (172.7) (74.47)*** |
| Pop. × Post | 273.0 (32.49)*** (19.10)*** | 257.8 (32.08)*** (23.13)*** | 274.1 (33.18)*** (19.27)*** | 264.1 (33.53)*** (22.91)*** | 272.1 (32.61)*** (19.19)*** | 258.5 (33.04)*** (22.79)*** |
| County Effects | Yes | Yes | Yes | Yes | Yes | Yes |
| Date Effects | Yes | Yes | Yes | Yes | Yes | Yes |
| Observations | 151945 | 151945 | 151945 | 151945 | 151945 | 151945 |

Notes: Odd columns report results from regression (5) and even columns report an IV version of this regression, where we instrument for $(Ads_i \times Post_t)$ and $(Ads_i \times Post_t \times W_i)$ using $(Treat_i \times Post_t \times W_i)$ and $(Treat_i \times Post_t)$. $W$ is an indicator for whether the value of a given county-level characteristic is below the median of that characteristic across counties in our sample. This characteristic is the 2016 Trump vote share in columns 1-2, the fraction of county residents with a college degree in columns 3-4, and the fraction of county residents who are white in columns 5-6. "***", "**", "*", and "+" indicate significance (from a two-tailed test) at the 0.01, 0.05, 0.10, and 0.20 levels. Standard errors, reported in parentheses below each estimate, are clustered at the state level in the first row and at the county level in the second row.

Table A.11: Measuring Vaccines in Rates

|  | Intent-to-Treat Effect | | Average Causal Response of 1,000 Ads | |
|---|---|---|---|---|
|  | (1) | (2) | (3) | (4) |
| Treat × Post | 0.570 | 0.563 | | |
|  | (0.386)+ | (0.371)+ | | |
|  | (0.437)+ | (0.437)+ | | |
|  | | | | |
| (Ads per 100) × Post | | | 0.0296 | 0.0291 |
|  | | | (0.0200)+ | (0.0192)+ |
|  | | | (0.0227)+ | (0.0226)+ |
|  | | | | |
| Pop. × Post | | -0.0234 | | -0.0270 |
|  | | (0.0561) | | (0.0580) |
|  | | (0.0134)* | | (0.0139)* |
| County Fixed Effects | Yes | Yes | Yes | Yes |
| Date Fixed Effects | Yes | Yes | Yes | Yes |
| Pop. × Post Dummy | Yes | | Yes | |
| Pop. × Date Dummies | | Yes | | Yes |
|  | | | | |
| Observations | 163856 | 163856 | 163856 | 163856 |

Notes: Table displays estimates from modifications of regressions (1) and (2) where the dependent variable is the total percent of the county population vaccinated at a given point in time and the treatment intensity is measured as the number of ads a county receives per 100 residents. "***", "**", "*", and "+" indicate significance (from a two-tailed test) at the 0.01, 0.05, 0.10, and 0.20 levels. Standard errors, reported in parentheses below each estimate, are clustered at the state level in the first row and at the county level in the second row.